\documentclass[12pt]{article}
\usepackage{epsfig,amssymb,amsmath,psfrag}

\catcode`\@=11
\DeclareMathAlphabet{\mathpzc}{OT1}{pzc}{m}{it}

\font\cmss=cmss12 
\def\1{\hbox{{1}\kern-.25em\hbox{l}}}
\def\bfZ{\relax{\hbox{\cmss Z\kern-.4em Z}}}

\def \be  {\begin{equation}}
\def \ee  {\end{equation}}
\def \ba  {\begin{eqnarray}}
\def \ea  {\end{eqnarray}}
\def \baa {\begin{eqnarray*}}
\def \eaa {\end{eqnarray*}}
\def \bb  {\begin {thebibliography} }
\def \eb  {\end{thebibliography}}
\def \lab #1 {\label{#1}}

\newcommand\re[1]{(\ref{#1})}

\def \matrix #1 {\left(\begin{array}{cc} #1 \end{array}\right)}

\def \tr {\mathop{\rm tr}\nolimits}

\newcommand{\as}{\ifmmode\alpha_{\rm s}\else{$\alpha_{\rm s}$}\fi}
\newcommand{\asbar}{\ifmmode\bar{\alpha}_{\rm s}\else{$\bar{\alpha}_{\rm s}$}\fi}

\newcommand{\ft}[2]{{\textstyle\frac{#1}{#2}}}

\font\cmss=cmss12 
\def\inbar{\,\vrule height1.5ex width.4pt depth0pt}
\def\IC{\relax\hbox{$\inbar\kern-.3em{\rm C}$}}
\def\IZ{\relax{\hbox{\cmss Z\kern-.4em Z}}}
\def\IR{{\hbox{{\rm I}\kern-.2em\hbox{\rm R}}}}

\def\IP{{\hbox{{\rm I}\kern-.2em\hbox{\rm P}}}}
\def\II{\hbox{{1}\kern-.25em\hbox{l}}}

\def\numberbysection{\@addtoreset{equation}{section}
                     \def\theequation{\thesection.\arabic{equation}}}
\numberbysection

\newbox\lett\newdimen\lheight\newdimen\lwidth
\def\ontop#1#2{\setbox\lett=\hbox{#2}\lheight\ht\lett
\multiply\lheight by 12 \divide\lheight by 10\relax%
\lwidth\wd\lett \multiply\lwidth by 8 \divide\lwidth by 10\relax #2\kern-\lwidth%
\raise\lheight\hbox{{$\scriptstyle #1$}}\kern.1ex}

\def\XXint#1#2#3{{\setbox0=\hbox{$#1{#2#3}{\int}$}
     \vcenter{\hbox{$#2#3$}}\kern-.5\wd0}}

\begin{document}

\begin{titlepage}

\thispagestyle{empty}

\vskip2cm

\centerline{\large \bf Strong coupling expansion of Baxter equation in $\mathcal{N}=4$ SYM}

\vspace{1cm}

\centerline{\sc A.V. Belitsky}

\vspace{10mm}

\centerline{\it Department of Physics, Arizona State University}
\centerline{\it Tempe, AZ 85287-1504, USA}

\vspace{1cm}

\centerline{\bf Abstract}

\vspace{5mm}

The anomalous dimensions of single-trace local Wilson operators with covariant derivatives
in maximally supersymmetric gauge theory are believed to be generated from a deformed
noncompact $sl(2)$ Baxter equation. We perform a systematic expansion of this equation at
strong coupling in the single-logarithmic limit of large conformal spin to overcome the
limitation of the asymptotic nature of the equation. The analysis is reduced to
Riemann-Hilbert problems for corresponding resolvents of Bethe roots in each order of the
quasiclassical expansion. We explicitly construct the resolvents in the lowest two orders
in strong coupling and find all local conserved charges of the underlying long-range spin
chain.

\end{titlepage}

\setcounter{footnote} 0

\newpage

\pagestyle{plain}
\setcounter{page} 1

\section{Introduction}

Until recently, the bulk of our knowledge about QCD or any quantum field
theory has been gained from methods relying on perturbative expansions
with respect to a small parameter. The lack of nonperturbative techniques
overcoming limitations of perturbative considerations was a major obstacle
in understanding strong coupling dynamics of gauge theories. The numerical
framework of lattice gauge theory partially resolves these problems, but
one still lacks reliable analytical methods. A concept stepping forward in
recent years as a panacea for these complications is integrability --- an
idea going back to the Liouville theorem, which states that a mechanical
system is integrable provided it has as many conserved quantities as degrees
of freedom such that all physical quantities can be calculated exactly.

In four-dimensional gauge theories integrable structures were first observed
in studies of multi-reggeon compound states \cite{Lip93,FadKor95}. Later analyses
of one-loop anomalous dimensions of the maximal-helicity Wilson operators in QCD
\cite{BraDerMan98,BraDerKorMan99,Bel99} revealed that their dilatation operator is
mapped in the multi-color limit into the Hamiltonian of the noncompact XXX
Heisenberg spin chain, and its eigenspectrum can be computed exactly by means of
the algebraic Bethe ansatz \cite{TakFad79}. Integrability observed in QCD anomalous
dimensions at one-loop order is a generic phenomenon of four-dimensional Yang-Mills
theories. Theories with supersymmetries obviously inherit integrability, although
the number of integrable sectors strongly depends on the particle content of the
models and is enhanced for theories with more supercharges \cite{BelDerKorMan04},
eventually encompassing all operators in maximally supersymmetric Yang-Mills
theory \cite{Lip98,MinZar03,BeiSta03,BeiKriSta03}.

The gauge/string duality for the $\mathcal{N} = 4$ super-Yang-Mills theory
\cite{Mal97} is of the weak/strong coupling nature and allows one to establish
exact correspondence between anomalous dimensions of composite Wilson operators
in gauge theory and energies of string excitations on the AdS$_5\times$S$^5$
background \cite{BerMalNas02,GubKlePol03}. It was demonstrated that the classical
string sigma models with anti-de Sitter space as a factor of the target space
possess an infinite set of integrals of motion and therefore are integrable
\cite{ManSurWad02}. On the gauge theory side, this result implies that the
all-order dilatation operator for Wilson operators should be integrable. A large
amount of evidence had been gathered to date about survival of integrability at
higher orders of perturbation theory culminating in a conjecture of asymptotic
all-order Bethe equations \cite{BeiSta05} depending on the Yang-Mills coupling
constant $g^2 = g^2_{\rm\scriptscriptstyle YM} N_c/(4 \pi^2)$ which incorporate
a phase factor encoding smooth interpolation from weak to strong coupling.

Recently we have formulated an alternative framework based on long-range Baxter
equations \cite{BelKorMul06,Bel06,Bel07} to eigenvalue problem of all-order
anomalous dimensions in maximally supersymmetric gauge theory. In our present
note we will address, generalizing earlier considerations of Ref.\ \cite{BelGorKor06},
the question of their systematic strong-coupling expansion restricting to the
noncompact $sl(2)$ sector and determine the spectrum of conserved charges of the
magnet in the single-logarithmic asymptotic in the conformal spin in the lowest
two orders of inverse-coupling series. One of these charges coincides with the
eigenvalue of the dilatation operator and in the large-spin limit defines the
cusp anomalous dimension --- an observable encoding the physics of soft-gluon
emission \cite{Col89,Kor88}. Recently a closed integral equation was formulated
for this quantity \cite{EdeSau06,Bel06,BeiEdeSta06} and its solution at strong
coupling to the lowest few orders was found numerically in Ref.\ \cite{BenBenKleSca06}
and analytically to the first \cite{AldAruBenEdeKle07,KotLip07,KosSerVol07,BecAngFor07}
and eventually at an arbitrary order in Ref.\ \cite{BasKorKot07}. These findings
match nicely with classical \cite{GubKlePol03,Kru06,BelGorKor06}, one
\cite{FroTse03,FroTirTse07} and two-loop calculations \cite{RoiTse07} in string
sigma model, and with the quantum string Bethe Ansatz analysis \cite{CasKri07}.

\section{Asymptotic Baxter equation}

The calculation of anomalous dimensions of arbitrary twist operators is a nontrivial
task even to one-loop order due to a large size of the mixing matrix. However, as we
pointed out above the problem can be overcome thanks to the hidden integrability
symmetry of the dilatation operator in $\mathcal{N} = 4$ super Yang-Mills theory in
the large-$N_c$ limit. In the closed holomorphic $sl(2)$ subsector of the theory,
spanned by twist-$L$ local Wilson operators built from scalar fields $X (0)$ and $N$
light-cone covariant derivatives (schematically)
\be
\label{WilsonOperators}
\mathcal{O}_{N,L} (0) = \tr \{ D_+^N X^L (0) \} \, ,
\ee
the one-loop mixing matrix is mapped into the Hamiltonian of the Heisenberg magnet of
length $L$ and spin $s = 1/2$ determined by the conformal spin of the field $X (0)$.
At higher orders of perturbation theory, the one-loop Bethe Ansatz equations \cite{BeiSta05}
or Baxter equations \cite{Bel06} allow for a consistent deformation with 't Hooft coupling
$g$ and yield spectra of anomalous dimensions in agreement with diagrammatic calculations.
They are plagued however by asymptotic nature, being inapplicable when the order of
perturbation theory exceeds the length $L$ of the local Wilson operator in question.

In our analyses we use the method of the Baxter equation \cite{Bax72} which proved to
be convenient in analyzing various semiclassical limits including the limit of large $L$
and $N$ at one-loop order \cite{Kor98,BraDerKorMan99,Bel99,BelGorKor06,GroKaz06}. The
method relies on the existence of an operator $\mathbb{Q} (u)$ which acts on the Hilbert
space of the chain and is diagonalized by all eigenstates of the magnet for an arbitrary
complex spectral parameter $u$. For discussion of the spectrum of integrals of motion of
the magnet it suffices to study just its eigenvalues $Q (u)$. Thus, the long-range $sl(2)$
Baxter equation is written for the polynomial
\be
Q (u) = \prod_{j = 1}^N (u - u_j)
\ee
with zeros determined by the Bethe roots $u_j = u_j (g)$ admitting an infinite series
expansion in 't Hooft coupling constant $g$,
\be
(x^+)^L {\rm e}^{\ft12 \left( \Delta_+ (x^+) - \Delta_- (x^-) \right)} Q (u + i)
+
(x^-)^L {\rm e}^{\ft12 \left( \Delta_- (x^-) - \Delta_+ (x^+) \right)} Q (u - i)
=
\tau (x) Q(u)
\, .
\ee
The deformation of the one-loop chain is partially encoded in the renormalized spectral
parameter $x \equiv x[u] = \ft12 \left( u + \sqrt{u^2 - g^2} \right)$ \cite{BeiSta05},
with adopted convention for $x^\pm = x [u \pm \ft{i}{2}]$. The right-hand side depends
on the transfer matrix with the two-dimensional auxiliary space which is a polynomial in
$x$ with expansion coefficients determined by the conserved charges of the chain\footnote{
Notice that compared to the equation in Ref.\ \cite{Bel06} we introduced a redefined
transfer matrix $\tau$, $t (x) = {\rm e}^{\ft12 \Delta_+ (x^+) + \ft12 \Delta_- (x^-)}
\tau (x)$, whose advantage compared to $t (x)$ is the absence of $x^{L - 1}$ terms and
thus corresponding charge in the transfer matrix \cite{BelKorMul06}.}.

The most nontrivial deviations from the nearest-neighbor magnet are accommodated in the
dressing factors $\sigma_\pm$ and $\Theta$ in the combination
\be
\Delta_\pm (x) = \sigma_\pm (x) - \Theta (x)
\, ,
\ee
with $\sigma_\pm$ responsible for the renormalization of the conformal spin of the Wilson
operators \re{WilsonOperators}
\be
\label{AllLoopSigma}
\sigma_\pm (x)
=
\int_{- 1}^1 \frac{dt}{\pi} \,
\frac{\ln Q \left( \pm \ft{i}{2} - g t \right)}{\sqrt{1 - t^2}}
\left(
1 - \frac{\sqrt{u^2 - g^2}}{u + g t}
\right)
\, ,
\ee
and $\Theta$ providing smooth matching of the weak- and strong-coupling expansion
\ba
\label{ThetaWeak}
\Theta (x)
\!\!\!&=&\!\!\!
g \int_{- 1}^1 \frac{d t}{\sqrt{1 - t^2}}
\ln \frac{Q ( - \ft{i}{2} - g t )}{Q ( + \ft{i}{2} - g t )}
\,
{-\!\!\!\!\!\!\int}_{-1}^1 ds \frac{\sqrt{1 - s^2}}{s - t}
\nonumber\\
&\times&
\int_{C_{[i, i \infty]}} \frac{d \kappa}{2 \pi i} \frac{1}{\sinh^2 (\pi \kappa)} \
\ln
\left( 1 + \frac{g^2}{4 x x [\kappa + g s]} \right)
\left( 1 - \frac{g^2}{4 x x [\kappa - g s]} \right)
\, .
\ea
This phase factor can be expressed as an infinite series expansion \cite{BeiEdeSta06}
\be
\label{ThetaStrong}
i \theta (x) \equiv \Theta (x^+) - \Theta (x^-)
=
i \sum_{j = 1}^N
\sum_{r = 2}^\infty \sum_{s = r + 1}^\infty
g^{r + s - 1}
c_{r,s} (g)
\left[
q_r (x) q_s (x_j)
-
q_r (x_j) q_s (x)
\right]
\ee
in terms of single-excitation charges on the spin chain
\be
q_r (x_j)
=
\frac{i}{r - 1}
\left(
\frac{1}{( x_j^+ )^{r - 1}}
-
\frac{1}{( x_j^- )^{r - 1}}
\right)
\, .
\ee
The latter define the local integrals of motion of the chain,
\be
\mathcal{Q}_r (g)
=
\sum_{j = 1}^N
q_r (x_j)
=
\frac{2 i}{r - 1} \left( - \frac{2}{g} \right)^{r - 2}
\int_{-1}^1 \frac{dt}{\pi} \, \sqrt{1 - t^2} \,
U_{r - 2} (t)
\left(
\ln  \frac{Q (+ \ft{i}{2} - g t)}{Q (- \ft{i}{2} - g t)}
\right)^\prime
\, ,
\ee
where $U_r (t)$ are the Chebyshev polynomials of the second kind. The anomalous
dimensions of the operators \re{WilsonOperators} are related to the Hamiltonian
$\mathcal{Q}_2$ via
\be
\label{AnomalousDimension}
\gamma (g)
= \frac{g^2}{2} \mathcal{Q}_2 (g)
\, .
\ee
At strong coupling, the expansion coefficients $c_{r,s} (g)$ were suggested in
Refs.\ \cite{AruFroSta03,HerLop05} and \cite{BeiEdeSta06} at leading, first subleading
and all orders, respectively. They represent analytic continuation of the phase
\re{ThetaWeak} suitable for perturbative analyses to strong coupling. The idea of
using the strong-coupling expansion of the scattering phase \re{ThetaStrong} has been
recently explored in Ref.\ \cite{CasKri07} within the framework of the asymptotic Bethe
Ansatz.

\section{Quasiclassical expansion}

As we pointed our earlier, due to asymptotic character of the Baxter equation it is
well defined only when the interaction range does not exceed the length of the chain.
Therefore, to study its strong-coupling expansion we have to consider a limit which
does not violate this requirement. It was established in Ref.\ \cite{BelGorKor06}, that
the anomalous dimensions of operators \re{WilsonOperators} for large conformal spin $N$
and length $L$ at strong coupling $g$ depend on the ``hidden'' parameter $\xi_{\rm str}
= g^2 \ln^2(N/L)/L^2$ and for $\xi_{\rm str} \gg 1$ the anomalous dimension becomes
insensitive to the twist $L$ of the operators thus circumventing limitations of the
asymptotic Baxter equation. In this asymptotic domain, the anomalous dimension scales
as $\sim \ln N$. Therefore, our goal is to develop the strong-coupling quasiclassical
expansion in the limit $g \gg 1$, $N \gg L \gg 1$ such that $\xi_{\rm str} \gg 1$.

To start with, let us construct a systematic expansion at strong coupling $g$. To this
end, we rescale the spectral parameters with a power of the coupling constant
\be
u = g \, \widehat{u}
\, , \qquad
x = g \, \widehat{x}
\, ,
\ee
such that the Baxter equation reads
\be
\widehat{x}^L \widehat{\tau} (\widehat{x})
= \left( \widehat{x}^- \right)^L
{\rm e}^{\ft12 \left( \widehat{\sigma}_- (\widehat{x}^-) - \widehat{\sigma}_+ (\widehat{x}^+) \right)
+ i \theta (\widehat{x})}
\frac{\widehat{Q} (\widehat{u} - \ft{i}{g})}{\widehat{Q} (\widehat{u})}
+
\left( \widehat{x}^+ \right)^L
{\rm e}^{\ft12 \left( \widehat{\sigma}_+ (\widehat{x}^+) - \widehat{\sigma}_- (\widehat{x}^-) \right)
- i \theta (\widehat{x})}
\frac{\widehat{Q} (\widehat{u} + \ft{i}{g})}{\widehat{Q} (\widehat{u})}
\, ,
\ee
where $\widehat{x}^\pm \equiv \widehat{x} [\widehat{u} \pm \ft{i}{2 g}]$ and the left-hand side
depends on the rescaled transfer matrix $\tau (x) = x^L \widehat{\tau} (\widehat{x})$. The magnon
phase $\theta (\widehat{x})$ is re-expressed in terms of the local charges
\be
\theta (\widehat{x})
=
\sum_{j = 1}^N \vartheta (\widehat{x}, \widehat{x}_j)
\equiv
g
\sum_{j = 1}^N
\sum_{r = 2}^\infty \sum_{s = r + 1}^\infty
c_{r, s} (g)
\left[
\widehat{q}_r (\widehat{x}) \widehat{q}_s (\widehat{x}_j)
-
\widehat{q}_r (\widehat{x}_j) \widehat{q}_s (\widehat{x})
\right]
\, ,
\ee
with the expansion coefficients $c_{r,s} (g)$ admitting the inverse-coupling expansion
\be
c_{r, s} (g)
=
c_{r, s}^{(0)} + \frac{1}{g} c_{r, s}^{(1)} + \frac{1}{g^2} c_{r, s}^{(2)} + \dots
\, .
\ee
The Baxter polynomial in the $\widehat{x}-$variable,
\be
\widehat{Q} (\widehat{u})
=
\prod_{j = 1}^N (\widehat{x} - \widehat{x}_j)
\prod_{j = 1}^N \left( 1 - \frac{1}{4 \widehat{x} \widehat{x}_j} \right)
\, ,
\ee
exhibits the double covering nature of the map $\widehat{u} \to \widehat{x}$, since
for each value of $\widehat{u}$ there are two corresponding values, $\widehat{x}$
and $1/(4 \widehat{x})$. While the second factor is irrelevant in perturbative
considerations, for finite coupling $g$ these zeros do contribute on equal footing.
However, as we will see below, their contribution vanishes from the Riemann-Hilbert
problems.

For our subsequent discussion, we introduce two resolvents
\be
S^\prime (\widehat{x}) = \eta \sum_{j = 1}^N \frac{1}{\widehat{x} - \widehat{x}_j}
\, , \qquad
G (\widehat{x})
= \eta \sum_{j = 1}^N
\frac{1}{(\widehat{x} - \widehat{x}_j) \left( 1 - \frac{1}{4 \widehat{x}_j^2} \right)}
\, ,
\ee
where $\eta = (N + \ft12 L)^{-1}$, which naturally emerge in the large-$g$ expansion
of the Baxter polynomials,
\ba
\ln \frac{\widehat{Q} (\widehat{u} \pm \ft{i}{g})}{\widehat{Q} (\widehat{u})}
\!\!\!&=&\!\!\!
\pm \frac{i}{g \eta}
\frac{
4 \widehat{x}^2 S^\prime (\widehat{x})
-
S^\prime (\frac{1}{4\, \widehat{x}})
}{4 \widehat{x}^2 - 1}
+
\frac{\eta}{2!}
\left( \frac{\pm i}{g \eta} \right)^2
\frac{
4 \widehat{x}^2 G^\prime (\widehat{x})
-
G^\prime (\frac{1}{4\, \widehat{x}})
}{4 \widehat{x}^2 - 1}
\\
&+&\!\!\!
\frac{\eta^2}{3!}
\left( \frac{\pm i}{g \eta} \right)^3
\left(
\frac{16 \, \widehat{x}^4 \, G^{\prime\prime} (\widehat{x})
+
G^{\prime\prime} (\frac{1}{4\, \widehat{x}})
}{\left( 4 \widehat{x}^2 - 1 \right)^2 }
-
\frac{
32 \, \widehat{x}^3 \,
[
G^{\prime} (\widehat{x})
-
G^{\prime} (\frac{1}{4\, \widehat{x}})
]
}{ \left( 4 \widehat{x}^2 - 1 \right)^3 }
+
\right)
+ \mathcal{O} (g^{-4})
\, . \nonumber
\ea
They are not independent though and are related to each other as follows
\be
\label{RelationGtoS}
G (\widehat{x}) = \frac{4 \, \widehat{x}^2 S^\prime (\widehat{x})}{4 \, \widehat{x}^2 - 1}
+ \frac{S^\prime (-\ft12)}{2 (2 \, \widehat{x} + 1)}
- \frac{S^\prime (\ft12)}{2 (2 \, \widehat{x} - 1)}
\, .
\ee
The phases also admit the inverse-coupling expansion
\ba
\ft12 \sigma_- (x^-) - \ft12 \sigma_+ (x^+)
\!\!\!&=&\!\!\!
\frac{i}{g \eta} \sigma_0 (\widehat{x}) + \frac{i}{g^3 \eta} \sigma_2 (\widehat{x})
+
\mathcal{O} (g^{-5})
\, ,
\nonumber\\
\vartheta (\widehat{x}, \widehat{y})
\!\!\!&=&\!\!\!
\frac{1}{g \eta} \vartheta_0 (\widehat{x}, \widehat{y})
+
\frac{1}{g^2 \eta} \vartheta_1 (\widehat{x}, \widehat{y})
+
\frac{1}{g^3 \eta} \vartheta_2 (\widehat{x}, \widehat{y})
+
\mathcal{O} (g^{-4})
\, ,
\ea
with
\ba
\vartheta_0 (\widehat{x}, \widehat{y})
\!\!\!&=&\!\!\!
\sum_{r = 2} \sum_{s = r + 1}
c_{r, s}^{(0)}
\left[
\widehat{q}_r^{(0)} (\widehat{x}) \widehat{q}_s^{(0)} (\widehat{y})
-
\widehat{q}_r^{(0)} (\widehat{y}) \widehat{q}_s^{(0)} (\widehat{x})
\right]
\, , \\
\vartheta_1 (\widehat{x}, \widehat{y})
\!\!\!&=&\!\!\!
\sum_{r = 2} \sum_{s = r + 1}
c_{r, s}^{(1)}
\left[
\widehat{q}_r^{(0)} (\widehat{x}) \widehat{q}_s^{(0)} (\widehat{y})
-
\widehat{q}_r^{(0)} (\widehat{y}) \widehat{q}_s^{(0)} (\widehat{x})
\right]
\, , \\
\vartheta_2 (\widehat{x}, \widehat{y})
\!\!\!&=&\!\!\!
\sum_{r = 2} \sum_{s = r + 1}
c_{r, s}^{(2)}
\left[
\widehat{q}_r^{(0)} (\widehat{x}) \widehat{q}_s^{(0)} (\widehat{y})
-
\widehat{q}_r^{(0)} (\widehat{y}) \widehat{q}_s^{(0)} (\widehat{x})
\right]
\\
&+&\!\!\!
\sum_{r = 2} \sum_{s = r + 1}
c_{r, s}^{(0)}
\left[
\widehat{q}_r^{(0)} (\widehat{x}) \widehat{q}_s^{(2)} (\widehat{y})
+
\widehat{q}_r^{(2)} (\widehat{x}) \widehat{q}_s^{(0)} (\widehat{y})
-
\widehat{q}_r^{(0)} (\widehat{y}) \widehat{q}_s^{(2)} (\widehat{x})
-
\widehat{q}_r^{(2)} (\widehat{y}) \widehat{q}_s^{(0)} (\widehat{x})
\right]
\, . \nonumber
\ea
These are expressed in terms of the strong-coupling coefficients
\cite{AruFroSta03,BeiTse05,HerLop05,BeiHerLop06}
\ba
c_{r, s}^{(0)} \!\!\!&=&\!\!\! \frac{1}{2^{2 r}} \delta_{r,s - 1}
\, , \\
c_{r, s}^{(1)} \!\!\!&=&\!\!\! - \frac{4}{\pi} \frac{1 - (-1)^{r + s}}{2^{r + s}}
\frac{(r - 1)(s - 1)}{(s + r - 2)(s - r)}
\, , \\
c_{r, s}^{(2)} \!\!\!&=&\!\!\!  \frac{1}{6} \frac{1 - (-1)^{r + s}}{2^{r + s}} (r - 1)(s - 1)
\, ,
\ea
and local integrals of motion
\be
\widehat{q}_r (\widehat{x})
=
\frac{1}{g} \widehat{q}_r^{(0)} (\widehat{x})
+
\frac{1}{g^3} \widehat{q}_r^{(2)} (\widehat{x})
+
\mathcal{O} (g^{- 5})
\, ,
\ee
where
\ba
\widehat{q}_r^{(0)} (\widehat{x})
\!\!\!&=&\!\!\!
\frac{4 \, \widehat{x}^{2 - r}}{4 \, \widehat{x}^2 - 1}
\, , \\
\widehat{q}_r^{(2)} (\widehat{x})
\!\!\!&=&\!\!\!
-
8 \, \widehat{x}^{4 - r}
\left(
\frac{r (r + 1)}{3 (4 \, \widehat{x}^2 - 1)^3}
+
\frac{2 (r + 1)}{(4 \, \widehat{x}^2 - 1)^4}
+
\frac{4}{(4 \, \widehat{x}^2 - 1)^5}
\right)
\, .
\ea
Finally, we assume that the resolvents and the transfer matrix admit systematic
expansion in inverse powers of the coupling
\ba
S (\widehat{x})
&=&
S_0 (\widehat{x}) + \frac{1}{g} S_1 (\widehat{x}) + \frac{1}{g^2} S_2 (\widehat{x}) + \dots
\, , \\
G (\widehat{x})
&=&
G_0 (\widehat{x}) + \frac{1}{g} G_1 (\widehat{x}) + \frac{1}{g^2} G_2 (\widehat{x}) + \dots
\, , \\
\tau (\widehat{x})
&=&
\tau_0 (\widehat{x}) + \frac{1}{g} \tau_1 (\widehat{x}) + \frac{1}{g^2} \tau_2 (\widehat{x}) + \dots
\, .
\ea
with each terms uniformly bounded. The coefficients $\tau_k (\widehat{x})$ are entire
functions in the complex plane $\widehat{x}$ with an essential singularity at $\widehat{x}
= 0$. Substituting these expansions into the Baxter equation and equating the coefficients
in front of powers of $g^{-1}$, find a set of equations for the resolvents.

\subsection{Leading order}

In explicit form, the leading contributions to the phase factors are
\ba
\sigma_0 (\widehat{x})
\!\!\!&=&\!\!\!
\frac{S_0^\prime (\ft12)}{2 (2 \, \widehat{x} - 1)}
- \frac{S_0^\prime (- \ft12)}{2 (2 \, \widehat{x} + 1)}
- \frac{2 S_0^\prime (\ft1{4 \, \widehat{x}})}{(4 \, \widehat{x}^2 - 1)}
\, , \\
\theta_0 (\widehat{x})
\!\!\!&=&\!\!\!
-
\frac{S_0^\prime (\ft12)}{2 (4 \widehat{x}^2 - 1)}
-
\frac{S_0^\prime (- \ft12)}{2 (4 \widehat{x}^2 - 1)}
+
\frac{S_0^\prime (\ft{1}{4 \widehat{x}})}{(4 \widehat{x}^2 - 1)}
\, ,
\ea
so that the zero-order quasiclassical form of the Baxter equation reads
\be
\label{LOBaxter}
\widehat{\tau}_0 (\widehat{x}) = 2 \cos p_0 (\widehat{x})
\, ,
\ee
in terms of the quasimomentum
\be
p_0 (\widehat{x})
=
\frac{1}{g \eta}
\left(
\frac{2 \beta \widehat{x}}{4 \widehat{x}^2 - 1}
+
\frac{
4 \widehat{x}^2 S_0^\prime (\widehat{x})
-
\widehat{x}
[ S_0^\prime (\ft12) - S_0^\prime (- \ft12) ]
}{4 \widehat{x}^2 - 1}
\right)
\, ,
\ee
where $\beta = L \eta$. From Eq.\ \re{LOBaxter} we find that $\sin p_0 (\widehat{x})$
is a double-valued function on the complex $\widehat{x}-$plane with the square-root
branching points $\widehat{x}_j$ obeying the condition $\tau_0 (\widehat{\xi}_j)=\pm 2$.
It becomes single-valued on a hyperelliptic Riemann surface defined by gluing together
two copies of the complex $\widehat{x}-$planes along the cuts running between the
branching points $\widehat{\xi}_j$, $S = [\widehat{\xi}_1, \widehat{\xi}_2] \cup
[\widehat{\xi}_3, \widehat{\xi}_4] \cup \dots$. In the quasiclassical limit, these
cuts accumulate the Bethe roots and correspond to regions of the allowed classical
motion of the system in separated variables.

From the single-valuedness of the transfer matrix, we can write down the Riemann-Hilbert
problem
\be
\label{LORH}
{\not\! p}_0 (\widehat{x}) = \pi m
\, ,
\ee
where here and below the principal value is defined as
\be
{\not\! p}_0 (\widehat{x})
\equiv \ft12
\left(
p_0 (\widehat{x} + i 0)
+
p_0 (\widehat{x} - i 0)
\right)
\, .
\ee
In this study we are interested in eigenstates possessing the minimal
possible energy for a given total spin $N$. For a given total spin $N$, this
trajectory is realized when all cuts but two in $S$ shrink into points and
the Bethe roots are located on two symmetric cuts on the real axis $[-a,-b]
\cup [b,a]$ most distant from the origin. From the point of view of separated
variables, this means that classically all but two collective degrees of
freedom are frozen and the classical motion is confined to the two intervals.
As a consequence, the complex spectral curve gets reduced to the genus one
curve \cite{BelGorKor06}. In Eq.\ \re{LORH}, the integer $m$ defines the
position of the interval $[b,a]$ inside $S$ and $m-1$ counts how many collapsed
cuts are situated to the right from the interval $[b,a]$ on the real axis. The
minimal value of the energy is achieved for $m = 1$.

Thus, to leading order of the semiclassical expansion the function $S_0'(\widehat{x})$
is determined by the above analytical properties and prescribed asymptotic behavior
at the origin and infinity
\be
S_0^\prime (\widehat{x} \to 0)
= \mathcal{\scriptstyle O} (\widehat{x})
\, , \qquad
S_0^\prime (\widehat{x} \to \infty)
=
\frac{\eta N}{\widehat{x}} + \mathcal{O} (\widehat{x}^{-2})
\, ,
\ee
for symmetric, $S^\prime_0 (- \widehat{x}) = - S^\prime_0 (\widehat{x})$, two-cut
solution. It reads
\be
\label{LOSolution}
S^\prime_0 (\widehat{x})
=
\frac{1}{4 \widehat{x}}
\oint_{C_1} \frac{d \widehat{z}}{2 \pi i}
\frac{W_0 (\widehat{z})}{\sqrt{(\widehat{z}^2 - a^2)(\widehat{z}^2 - b^2)}}
\left\{
\frac{1 - 4 \widehat{x}^2}{\widehat{z}^2 - \widehat{x}^2}
\sqrt{(\widehat{x}^2 - a^2)(\widehat{x}^2 - b^2)}
+
\frac{a b}{4 \widehat{z}^2}
\right\}
\, ,
\ee
where the contour $C_1$ goes around the cut $[b, a]$ in counterclockwise direction and
\be
W_0 (\widehat{z}) = 2 (g \eta) \pi m - \frac{4 \beta \widehat{z}}{4 \widehat{z}^2 - 1}
\, .
\ee
From the asymptotic behavior of $S_0^\prime (\widehat{x} \to \infty)$ one deduces conditions
on the end-points of the cuts
\ba
\label{Conditions}
&&
\oint_{C_1} \frac{d \widehat{z}}{2 \pi i}
\frac{W_0 (z)}{\sqrt{(\widehat{z}^2 - a^2)(\widehat{z}^2 - b^2)}}
= 0
\, , \\
&&
\oint_{C_1} \frac{d \widehat{z}}{2 \pi i}
\frac{W_0 (z)}{\sqrt{(\widehat{z}^2 - a^2)(\widehat{z}^2 - b^2)}}
\left(
\widehat{z}^2 + \frac{a b}{4 \widehat{z}^2}
\right)
= \eta N
\, . \nonumber
\ea
As follows from these conditions, the branching points of the curve, $\pm b$ and $\pm a$,
depend on the ratio $L/N$ and the coupling constant $g$. As was demonstrated in Ref.\
\cite{BelGorKor06}, the single-logarithmic asymptotics emerges for the configuration when
$b$ approaches its minimal value $\ft12$ so that the inner boundaries of two cuts $[-a_,-b]$
and $[b,a]$ coincide with the position of poles at $|\widehat{x}| = \ft12$ and the outer
points $\pm a$ run away to infinity:
\be
a \to \infty
\, , \qquad
b \to \ft12 + \varepsilon
\, .
\ee
From Eqs.\ \re{Conditions} we find the following parametric dependence of $a$ and $b$
on $N$, $L$ and $g$
\be
\label{EndPointsSingleLog}
\frac{1}{\sqrt{\varepsilon}} = \frac{4 m g}{L} \ln (4 a)
\, , \qquad
a = \frac{N}{2 m g}
\, ,
\ee
with the parameter $\varepsilon$ related to the string parameter introduced in Ref.\
\cite{BelGorKor06} as follows $\varepsilon \sim \xi^{- 2}_{\rm\scriptstyle str}$.
The analysis of the solution \re{LOSolution} yields in the single-logarithmic regime
\be
S^\prime_0 (\widehat{x})
= - \frac{\beta}{2 \sqrt{\varepsilon}}
\frac{i \sqrt{\widehat{x}^2 - b^2} + b}{\widehat{x} }
\, ,
\ee
or in terms of the resolvent $G_0$,
\be
\label{G0analytical}
G_0 (\widehat{x}) = - \frac{2 i \beta}{\sqrt{\varepsilon}}
\frac{\widehat{x} \sqrt{\widehat{x}^2 - b^2}}{ 4 \widehat{x}^2 - 1}
\, ,
\ee
which agrees with Ref.\ \cite{CasKri07}.

\subsection{Next-to-leading order}

Collecting the terms at the order $1/g$ in the expansion of the Baxter equation, we get
the equation for the next-to-leading correction to the resolvent
\be
G_1 (\widehat{x})
=
\frac{g \eta}{2}
\frac{p_0^\prime \left( \ft{1}{4 \widehat{x}} \right)
-
4 \widehat{x}^2 p_0^\prime (\widehat{x})
}{4 \widehat{x}^2 - 1}
\cot p_0 (\widehat{x})
+
\theta^{(1)}_0 (\widehat{x})
+
\frac{2 \beta \widehat{x}^2 (4 \widehat{x}^2 + 1)}{(4 \widehat{x}^2 - 1)^3}
\cot p_0 (\widehat{x})
-
\frac{g \eta}{2} \frac{\tau_1 (\widehat{x})}{\sin p_0 (\widehat{x})}
\, .
\ee
The Riemann-Hilbert problem for the subleading resolvent follows from this assuming
that $G_1 (\widehat{x})$ possesses the same analytic properties as the leading
$G_0 (\widehat{x})$. Therefore, since $p^\prime_0 (\widehat{x})$ and $\sin p_0
(\widehat{x})$ change sign across the cuts as a consequence of the leading order
Riemann-Hilbert problems \re{LORH}, we immediately find
\be
\label{NLORH}
{\not\! G}_1 (\widehat{x}) = - \frac{g \eta}{2} \frac{4 \widehat{x}^2}{4 \widehat{x}^2 - 1}
p_0^\prime (\widehat{x} + i 0) \cot p_0 (\widehat{x} + i 0)
+
\theta_0^{(1)} (\widehat{x})
\, ,
\ee
where we used the fact that while $S^\prime (\widehat{x})$ (and $G (\widehat{x})$) is
discontinuous on the cuts $x \in [-a, -b] \cup [b, a]$, $S^\prime (\ft{1}{4\widehat{x}})$
(and $G (\ft{1}{4\widehat{x}})$) for the reflected argument is continuous. Here
\be
\theta^{(1)}_0 (\widehat{x})
=
\frac{1}{4 \widehat{x}^2 - 1} \frac{2}{\pi}
\oint_{C} \frac{d \widehat{z}}{2 \pi i}
\frac{S^\prime_0 (\widehat{z})}{4 \widehat{z}^2 - 1}
\vartheta_1 (\widehat{x}, \widehat{z})
\, ,
\ee
with the contour $C$ wrapping around the cuts $[-a, -b] \cup [b, a]$ and where
\ba
\vartheta_1 (\widehat{x}, \widehat{z})
\!\!\!&=&\!\!\!
- 4 \sum_{r = 2}^\infty \sum_{m = 0}^\infty
\frac{(r - 1) (2m + r)}{(2m + 2r - 1)(2m + 1)}
\left[
(2 \widehat{x})^{2 - r} (2 \widehat{z})^{2m + r - 1}
-
(2 \widehat{z})^{2 - r} (2 \widehat{x})^{2m + r - 1}
\right]
\nonumber\\
&=&\!\!\!
(4 \widehat{x} \widehat{z})^2
\left\{
\frac{1}{(4 \widehat{x} \widehat{z} - 1) (\widehat{x} - \widehat{z})}
+
\left(
\frac{1}{(4 \widehat{x} \widehat{z} - 1)^2}
+
\frac{1}{4 (\widehat{x} - \widehat{z})^2}
\right)
\ln \frac{(2 \widehat{x} + 1)(2 \widehat{z} - 1)}{(2 \widehat{x} - 1)(2 \widehat{z} + 1)}
\right\}
\, .
\ea
The solution to the Riemann-Hilbert problem \re{NLORH} is
\be
G_1 (\widehat{x}) = \widehat{x} \oint_{C_1} \frac{d \widehat{z}}{2 \pi i}
W_1 (\widehat{z})
\left\{
\frac{1}{\widehat{x}^2 - \widehat{z}^2}
+
\frac{1}{1 - 4 a b} \frac{1}{\widehat{z}^2}
\right\}
\frac{\sqrt{(\widehat{z}^2 - a^2)(\widehat{z}^2 - b^2)}
}{
\sqrt{(\widehat{x}^2 - a^2)(\widehat{x}^2 - b^2)}}
\, ,
\ee
where
\be
W_1 (\widehat{x})
=
2 \theta_0^{(1)} (\widehat{x})
-
g \eta \frac{4 \widehat{x}^2}{4 \widehat{x}^2 - 1}
p_0^\prime (\widehat{x} + i 0) \cot p_0 (\widehat{x} + i 0)
\, ,
\ee
in agreement with Ref.\ \cite{CasKri07}. A tedious calculation yields the
next-to-leading resolvent in the single-logarithmic asymptotics
\ba
\label{G1analytical}
G_1 (\widehat{x})
\!\!\!&=&\!\!\!
\frac{i \beta}{8 \pi \sqrt{\varepsilon}}
\bigg\{
-
\frac{2 \widehat{x} (2 b - 1) (12 \widehat{x}^2 + 1)}{(4 \widehat{x}^2 - 1)^2 \sqrt{\widehat{x}^2 - b^2}}
-
\frac{64 \widehat{x}^2 (4 \widehat{x}^2 + 1)\sqrt{4 b^2 - 1}}{(4 \widehat{x}^2 - 1)^3}
\cot^{- 1} \left( \widehat{x} \frac{\sqrt{4 b^2 - 1}}{\sqrt{\widehat{x}^2 - b^2}} \right)
\nonumber\\
&-&\!\!\!
\frac{128 \widehat{x}^3 [1 - 2 b^2 (4 \widehat{x}^2 + 1)]}{(4 \widehat{x}^2 - 1)^3 \sqrt{16 \widehat{x}^2 b^2 - 1}}
\cot^{- 1} \left( \frac{\sqrt{16 \widehat{x}^2 b^2 - 1}}{2 \sqrt{\widehat{x}^2 - b^2}} \right)
-
\frac{2 \widehat{x} (4 b^2 - 1)}{(4 \widehat{x}^2 - 1) \sqrt{\widehat{x}^2 - b^2}} \ln \left( 1 - \frac{1}{4 b^2} \right)
\nonumber\\
&+&\!\!\!
\frac{
\widehat{x} \left[ 2 (16 \widehat{x}^4 - 1) - (4 b^2 - 1) (1 + 24 \widehat{x}^2 + 16 \widehat{x}^4) \right]
}{
(4 \widehat{x}^2 - 1)^3 \sqrt{\widehat{x}^2 - b^2}
}
\ln \left[ \left( 1 + \ft{1}{2 b} \right)^2 \left( 1 + \ft{1}{4 b^2} \right) \right]
\nonumber\\
&+&\!\!\!
\frac{32 \widehat{x}^2 [ 2 \widehat{x}^2 -  b^2(4 \widehat{x}^2 + 1)]}{(4 \widehat{x}^2 - 1)^3 \sqrt{\widehat{x}^2 - b^2}}
\ln \frac{(\widehat{x} - b)(2 \widehat{x} + 1)}{(\widehat{x} + b)(2 \widehat{x} - 1)}
\bigg\}
\, .
\ea
While, the resolvent $S_1^\prime (\widehat{x})$ can be found from it making use of the
relation \re{RelationGtoS}.

\section{Local integrals of motion}

For the determination of the local integrals of motion $\mathcal{Q}_r (g)$, it suffices
to use reduced resolvents with inner end points of the cuts collided with the poles at
$|\widehat{x}| = \ft12$. They immediately follow from Eqs.\ \re{G0analytical} and
\re{G1analytical},
\ba
G_0 (\widehat{x})
\!\!\!&=&\!\!\!
- \frac{i \beta}{\sqrt{\varepsilon}}
\frac{\widehat{x}}{\sqrt{4 \widehat{x}^2 - 1}}
\, , \\
G_1 (\widehat{x})
\!\!\!&=&\!\!\!
\frac{i \beta}{2 \pi \sqrt{\varepsilon}}
\frac{\widehat{x}
\left[
4 \pi \widehat{x}^2 + 3 (4 \widehat{x}^2 + 1) \ln 2
\right]
}{\sqrt{(4 \widehat{x}^2 - 1)^5}}
\, .
\ea
Since the discontinuity of the resolvent generates the distribution of Bethe roots, we
conclude from the subleading order in the strong-coupling expansion that the enhanced
singularity at $|\widehat{x}| = \ft12$ implies stronger accumulation of Bethe roots around
these poles. This is to be contrasted with weak-coupling distributions where the poles do
not visible. Then the strong-coupling expansion of the charges
\be
\mathcal{Q}_r (g)
=
\mathcal{Q}^{(0)}_r + \frac{1}{g} \mathcal{Q}^{(1)}_r + \mathcal{O} (g^{-2})
\ee
is obtained by evaluating residues of the resolvents at $\widehat{x} = 0$,
\be
\mathcal{Q}^{(k)}_r (g)
=
- \frac{1}{g^r \eta}
\oint_{|\widehat{x}| < \delta } \frac{d \widehat{x}}{2 \pi i} \,
\widehat{x}^{- r} G_k (\widehat{x})
\, ,
\ee
with $k = 1, 2$. Explicitly, they read
\ba
\mathcal{Q}^{(0)}_r
\!\!\!&=&\!\!\!
\frac{L}{g^r \sqrt{\varepsilon}}
\frac{\Gamma (r - 1)}{\Gamma^2 (\ft{r}{2})}
\, , \\
\mathcal{Q}^{(1)}_r
\!\!\!&=&\!\!\!
- \frac{L}{g^r \sqrt{\varepsilon}}
\frac{\Gamma (r)}{\Gamma^2 (\ft{r}{2})}
\left[
\frac{r - 2}{6} + (2 r - 1) \frac{\ln 2}{2 \pi}
\right]
\, .
\ea
The eigenvalue of the Hamiltonian $\mathcal{Q}_2$ agrees with previous calculations
of the cusp anomalous dimension at strong coupling
\cite{AldAruBenEdeKle07,KotLip07,KosSerVol07,BecAngFor07,BasKorKot07,GubKlePol03,Kru06,BelGorKor06,FroTse03,FroTirTse07,CasKri07}.

\section{Conclusions}

In the present paper we have suggested a systematic quasiclassical expansion
of all-order Baxter equation in the noncompact $sl(2)$ sector of the maximally
supersymmetric Yang-Mill theory at strong coupling. We have focused on the
single-logarithmic asymptotics in the conformal spin corresponding to leading
order contribution in the parameter $\varepsilon$, see Eq.\ \re{EndPointsSingleLog}.
We found the generating function for all local conserved charges $\mathcal{Q}_r$
in the first two orders of the inverse-coupling expansion. For $r = 2$, we reproduce
cusp anomalous dimension at strong coupling found by other techniques.

Our approach can immediately be used to compute corrections order-by-order in
$\varepsilon-$expansion. It can further be employed to find subsequent terms in
the inverse-coupling expansion by computing higher-order terms to the resolvent.
For instance, at next-to-next-to-leading order, we find the following Riemann-Hilbert
problem
\be
{\not\! G}_2 (\widehat{x})
=
-
\frac{\widehat{x}^2}{4 \widehat{x}^2 - 1}
\left[ \left( G_1 (\widehat{x} + i 0)
- W_1 (\widehat{x})
\right) \cot p_0 (\widehat{x}) \right]^\prime
+
\theta^{(1)}_1 (\widehat{x})
+
g_2 (\widehat{x})
\, , \nonumber
\ee
where $g_2 (\widehat{x})$ is
\ba
g_2 (\widehat{x})
=
\!\!\!&-&\!\!\!
\frac{8 \beta \widehat{x}^3 (1 + 16 \widehat{x}^2 + 16 \widehat{x}^4)}{(4 \widehat{x}^2 - 1)^5}
- \frac{1}{1536} \frac{\widehat{x} S^{(5)}_0 (\ft12)}{(4 \widehat{x}^2 - 1)}
- \frac{5}{384} \frac{\widehat{x} S^{(4)}_0 (\ft12)}{(4 \widehat{x}^2 - 1)}
\nonumber\\
&-&\!\!\! \frac{1}{384}
\frac{
\widehat{x} (21 - 296 \widehat{x}^2 + 336 \widehat{x}^4)
S^{\prime\prime\prime}_0 (\ft12)}{(4 \widehat{x}^2 - 1)^3}
- \frac{1}{64}
\frac{
\widehat{x} ( (1 + 4 \widehat{x}^2)^2 - 144 \widehat{x}^2)
S^{\prime\prime}_0 (\ft12)}{(4 \widehat{x}^2 - 1)^3}
\nonumber\\
&+&\!\!\!
\frac{1}{96}
\frac{
\widehat{x} ( 3 + 104 \widehat{x}^2 + 48 \widehat{x}^4)
S^{\prime}_0 (\ft12)}{(4 \widehat{x}^2 - 1)^3}
\, ,
\ea
and
\be
\theta^{(1)}_1 (\widehat{x})
=
\frac{1}{4 \widehat{x}^2 - 1} \frac{2}{\pi}
\oint_{C} \frac{d \widehat{z}}{2 \pi i}
\frac{S^\prime_1 (\widehat{z})}{4 \widehat{z}^2 - 1}
\vartheta_1 (\widehat{x}, \widehat{z})
\, .
\ee
Its analysis deserves, however, a separate study.

\vspace{0.5cm}

We would like to thank Gregory Korchemsky and Arkady Tseytlin for instructive
communications and very useful comments on the manuscript. This work was supported
by the U.S.\ National Science Foundation under grant no.\ PHY-0456520.

\end{document}